\documentclass[conference]{IEEEtran}
\IEEEoverridecommandlockouts

\usepackage{cite}
\usepackage{amsmath,amssymb,amsfonts}
\usepackage{algorithmic}
\usepackage{graphicx}
\usepackage{textcomp}
\usepackage{xcolor}
\def\BibTeX{{\rm B\kern-.05em{\sc i\kern-.025em b}\kern-.08em
    T\kern-.1667em\lower.7ex\hbox{E}\kern-.125emX}}

\usepackage{atbegshi,picture}
\AtBeginShipoutNext{\AtBeginShipoutUpperLeft{%
  \put(\dimexpr\paperwidth-5.2cm\relax,-1.5cm){\makebox[0pt][r]{Accepted manuscript 11th International IEEE/EMBS Conference on Neural Engineering (NER) 2023}}%
}}

\begin{document}

\title{Towards AI-controlled FES-restoration of arm movements: Controlling for progressive muscular fatigue with Gaussian state-space models}
\author{Nat Wannawas$^1$ \&
        A. Aldo Faisal$^{1,2}$
\thanks{$^{1}$Brain \& Behaviour Lab, Imperial College London, London SW7 2AZ, United Kingdom. (email:nat.wannawas18@imperial.ac.uk). $^2$ Chair in Digital Health \& Data Science, University of Bayreuth, Bayreuth, Germany. (aldo.faisal@imperial.ac.uk).
NW acknowledges his support by the Royal Thai Government Scholarship. AAF acknowledges his support by UKRI Turing AI Fellowship (EP/V025449/1).}
}
\maketitle

\begin{abstract}
    Reaching disability limits an individual's ability in performing daily tasks. Surface Functional Electrical Stimulation (FES) offers a non-invasive solution to restore the lost abilities. However, inducing desired movements using FES is still an open engineering problem. This problem is accentuated by the complexities of human arms' neuromechanics and the variations across individuals. Reinforcement Learning (RL) emerges as a promising approach to govern customised control rules for different subjects and settings. Yet, one remaining challenge of using RL to control FES is unobservable muscle fatigue that progressively changes as an unknown function of the stimulation, breaking the Markovian assumption of RL.
    
    In this work, we present a method to address the unobservable muscle fatigue issue, allowing our RL controller to achieve higher control performances. Our method is based on a Gaussian State-Space Model (GSSM) that utilizes recurrent neural networks to learn Markovian state-spaces from partial observations. The GSSM is used as a filter that converts the observations into the state-space representation for RL to preserve the Markovian assumption. Here, we start with presenting the modification of the original GSSM to address an overconfident issue. We then present the interaction between RL and the modified GSSM, followed by the setup for FES control learning. We test our RL-GSSM system on a planar reaching setting in simulation using a detailed neuromechanical model and show that the GSSM can help RL maintain its control performance against the fatigue.
\end{abstract}

\section{Introduction}
Yearly, strokes and spinal cord injuries have left individuals around the world with paralysis. Upper body paralysis, one of the most commonly found following incidents, causes the dysfunction of arm movements and severely affects the individuals' abilities in performing daily tasks. Functional Electrical Stimulation (FES), a technique that uses electrical signals to induce muscle contraction, offers a solution for restoring the movements. Yet, controlling FES to induce desired movements is challenging. One challenge is that each individual requires customised stimulation to induce a certain movement. This causes difficulties in designing a control method that works across different individuals without intensive, manual configurations. Another challenge is that the muscle's responses to the FES change over time because of muscular fatigue. Since the fatigue level can not be directly measured, it is difficult for a controller to maintain its performance over extended periods.

Several methods that can automatically find customised stimulation have been investigated. One of those is Reinforcement Learning (RL), a machine learning algorithm with a learning agent (RL agent) that learns to control an environment through interaction. The successes of RL in controlling body movements have been presented in several scenarios: cycling \cite{Wannawas2021}, walking \cite{Anand2019}, arm movements \cite{Thomas2008,Jagodnik2016,Wolf2020,Wannawas2022,Abreu2022}. Additionally, \cite{Abreu2022} shows that RL can deal with fatigue to a certain degree; yet, the performance drop is still inevitable in many cases.

Different approaches have been employed to deal with muscular fatigue. A widely used approach is to record electromyogram (EMG) or mechanomyogram (MMG) from which muscle force can be estimated \cite{Woods2018,Islam2018,Naeem2020,Krueger2020}. Although this approach could be straightforward, the successes are, currently, limited to a few types of movements such as knee extension \cite{Naeem2020,Krueger2020} and cycling \cite{Woods2018,Islam2018}. Additionally, it requires sensors which can be difficult to set up. Approaches that exploit the periodic nature of the movements such as walking are used in \cite{Ama2014,Ha2016}. However, these may not be suitable to be used in controlling arbitrary arm movements. Our previous work \cite{Wannawas2022} explores an approach that does not use dedicated sensors and can be applied to arbitrary movements. The approach uses a recurrent neural network (RNN) to encode the history of observations and provide additional information to the RL agent. This strategy can control arbitrary single-joint movements in the real world. However, its capability in multiple-joint cases is limited.

In this work, we present an AI-based system for controlling FES that can induce arbitrary desired movements and can maintain performances under progressive muscular fatigue. Our system uses the combination of an RNN-based Gaussian state-space model (GSSM) that learns Markovian state representations and RL that learns the control policies on the representation spaces. In simple terms, the GSSM here functions as a filter that provides insight information of the systems' states to the RL agents. Compared to our previous work \cite{Wannawas2022}, this system is more powerful and capable of learning the complex dynamics of multiple-joint movements. Additionally, it produces probabilistic transition functions that can be useful, for example, for model-based RL.

We present the details of our RL-GSSM system and the setup for controlling arbitrary movements in the \emph{Methods} section. We also provide the modification of the original GSSM \cite{Krishnan2017} to address overconfident issues. We demonstrate our RL-GSSM in a planar arbitrary reaching setting using a neuromechanical model and show that our RL-GSSM can maintain its control performance under the progressive fatigue.

\begin{figure*}[ht!]
    \begin{center}
    \includegraphics[width=0.96\linewidth]{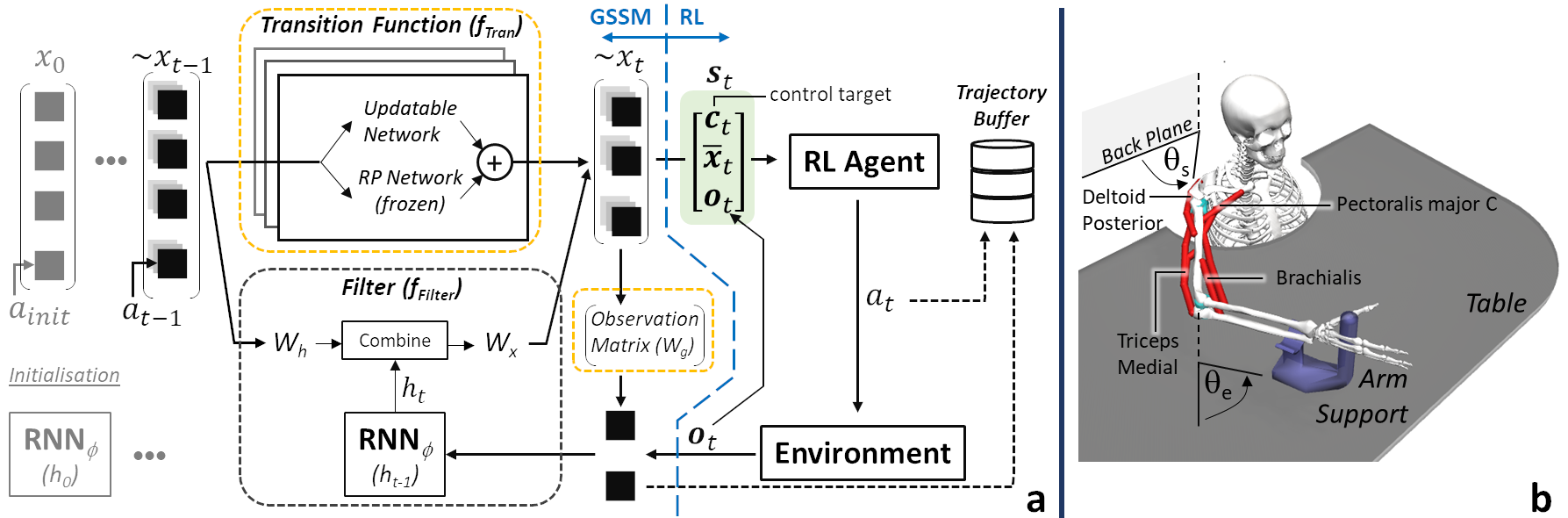}
    \vspace{-5pt}
    \caption{(a) Diagram showing the overview of our RL-GSSM system. The dash blue line splits RL and GSSM. The GSSM's parts in yellow boxes are excluded during the interaction phase. This phase starts with the initialisation (on the left) and evolves as follows. At the time step $t$, The previous action $\textbf{a}_{t-1}$ are appended to the state-representations of the previous time step $\textbf{x}_{t-1}$. The filter then combines the appended vector with the incoming observation $\textbf{o}_t$ and samples the state-representations of the current time step $\textbf{x}_t$. The average of $\textbf{x}_t$, denoted as $\boldsymbol{\bar{x}}_t$, is concatenated with $\textbf{o}_t$ and a control target $\textbf{c}_t$ and become an RL's state vector $\textbf{s}_t$. The interaction data are stored in Trajectory Buffer. (b) Detailed neuromechanical model of planar movement built using OpenSim.}
    \label{fig:diagram}
    \end{center}
    \vspace{-12pt}
\end{figure*}

\section{Methods}

\subsection{Gaussian State-Space Models (GSSM)}
Here, GSSM functions as a filter that converts an observable environment's state vector ($\textbf{o}_t$) into a state-representation vector ($\textbf{x}_t$) which contains the information of the system's hidden states. Our GSSM is based on \cite{Krishnan2017} whose main components are an RNN-based filter ($f_{\mbox{Filter}}$) and a transition function ($f_{\mbox{Tran}}$). The filter converts $\textbf{o}_t$ into $\textbf{x}_t$ through a process described as follows. The process starts at the zeroth time step ($t=0$) with the initialisation of the RNN's hidden states ($\textbf{h}_0$) and state representations ($\textbf{x}_0$). $\textbf{x}_0$ is then concatenated with the initial action vector $\textbf{a}_{init}$ and is passed through $W_s$, a small multilayer perceptron (MLP). This step is mathematically expressed as $\textbf{h}_{x,t=0}=W_s(\textbf{x}_0;\textbf{a}_0]^T)$. Meanwhile, the RNN observes the environment's states $\textbf{o}_0$ and updates its hidden state to $\textbf{h}_{t=1}$. $\textbf{h}_{x,t=0}$ and $\textbf{h}_{t=1}$ are then combined as $\textbf{h}_{c,t=1}=\frac{1}{2}\tanh(\textbf{h}_{x,t=0}+\textbf{h}_{t=1})$. Next, $\textbf{h}_{c,t=1}$ is passed through $W_x$ which is an MLP that outputs the distribution of $\textbf{x}_t$. The following time steps repeat this process but start with the sampled $\textbf{x}_t$ and actual actions $\textbf{a}_t$. The trajectory of $\textbf{x}_t$, denoted as $\textbf{x}_{0:T}$, is obtained by repeating this process through the whole trajectory of observations $\textbf{o}_{0:T}$. For future notation, RNN, $W_h$, and $W_x$ are collectively referred as $f_{\mbox{Filter}}$.

The GSSM is trained using the trajectory of the observations ($\textbf{o}_{0:T}$) as follows. The training process starts with using $f_{\mbox{Filter}}$ to sample $\textbf{x}_{0:T}$ corresponding to $\textbf{o}_{0:T}$. Next, we reconstruct the observations by passing the sampled $\textbf{x}_{0:T}$ through the observation mapping function $W_g$, expressed as $\textbf{k}_{0:T}=W_g(\textbf{x}_{0:T})$. The parameters of $f_{\mbox{Filter}}$ are optimised through gradient descent to minimise the following loss functions. The first loss function is the likelihood between $\textbf{k}_{0:T}$ and $\textbf{o}_{0:T}$, expressed as $l_{lik}=\sum_{t=1}^Tp(\textbf{o}_t|\boldsymbol{\mu}_{k,t},\boldsymbol{\Sigma}_{k,t})$, where $\boldsymbol{\mu}_{k,t}$ and $\Sigma_{k,t}$ are the mean and covariance of the reconstructed observations, respectively. The second loss function is the KL divergence between the $\textbf{x}_{0:T}$ distribution sampled by $f_{\mbox{Filter}}$ and those predicted by $f_{\mbox{Tran}}$, expressed as
\begin{equation*}
 l_{D_{KL}}=\sum_{t=2}^TD_{KL}[f_{\mbox{Filter}}(\textbf{x}_{t-1},\textbf{o}_{0:t})||f_{\mbox{Tran}}(\textbf{x}_{t-1})].
\end{equation*}
Intuitively, this loss function encourages the filter-generated distribution of $\textbf{x}_t$, $p_f(\textbf{x}_t)$, to have a Markovian structure, i.e, $p_f(\textbf{x}_t|\textbf{x}_{t-1},\textbf{o}_{0:t}) = p(\textbf{x}_t|\textbf{x}_{t-1})$. Note that the observation history $\textbf{o}_{0:t-1}$ is encoded in the RNN's hidden states.

In the original model \cite{Krishnan2017}, $f_{\mbox{Tran}}$ is represented by a neural network that directly outputs the means and variances of $\textbf{x}_t$. This network, in some cases, produces overconfidence in the learned transition function. To mitigate this issue, we replace that network with the ensemble of neural networks with randomised prior functions (RP-Ensemble) \cite{Osband2018}. The predictive means and variances are computed by fitting Gaussian distributions to the ensemble's outputs.

\subsection{Generic RL-GSSM for controlling arbitrary movements}
Reinforcement Learning (RL) learns a task through reward signals collected from interactions with an environment. The interactions occur in a discrete-time fashion, starting with the agent observing the environment's state $\textbf{s}_t$ and selecting an action $\textbf{a}_t$ based on its policy $\pi$. The action causes the environment to be in a new state $\textbf{s}_{t+1}$. The agent then receives an immediate reward $r_t$ and observes the new state. This interaction experience is collected as a tuple $(\textbf{s}_t, \textbf{a}_t, r_t, \textbf{s}_{t+1})$ which is stored in a replay buffer $\mathcal{D}$. This tuple is used to learn an optimal policy $\pi^*$ that maximises a return $R$--the sum of discounted immediate rewards.

The introduction of GSSM into the system causes few changes in the learning process. To avoid confusing notation, we hereafter use $\textbf{s}_t$ to denote RL state vectors. Fig.\ref{fig:diagram}a shows the overview diagram of our RL-GSSM system. The system has two phases--interaction and updating phases--described as follows. At each time step in the interaction phase, $f_{\mbox{Filter}}$ observes $\textbf{o}_t$, updates the RNN's hidden states, and generates state-representations $\textbf{x}_t$. The agent then selects an action $\textbf{a}_t$ based on $\textbf{s}_t=[\textbf{o}_t; \textbf{x}_t; \textbf{c}_t]^T$, where $\textbf{c}_t$ is a control target at time $t$. The action affects the environment, the system moves into the next time step, and the process repeats. The interactions are stored as $([\textbf{o}_t;\textbf{c}_t]^T, \textbf{a}_t, r_t, [\textbf{o}_{t+1};\textbf{c}_{t+1}]^T)$ in a Trajectory Buffer.

The updating phase begins with drawing sampled trajectories ($\tilde{\textbf{o}}_{0:T}$) from the Trajectory Buffer and using them to update the GSSM. After that, the updated $f_{\mbox{Filter}}$ is used to generate new trajectories of $\textbf{s}_t$ corresponding to $\tilde{\textbf{o}}_{0:t}$. The new $\textbf{s}_t$ trajectories are then converted into new RL experience tuples stored in a typical Replay Buffer. After that, the RL agent is updated following a conventional method.

\subsection{RL-GSSM setup for controlling planar movements} 
\emph{The environment} here is a neuromechanical model built using OpenSim. The model has a human arm placed on an arm support that moves with low friction on a table (Fig.\ref{fig:diagram}b). The model has 6 muscles; 4 muscles--\emph{Brachialis}, \emph{Triceps medial}, \emph{Deltoid posterior}, and \emph{Pectoralis major c}--are stimulated. The muscles are fatigued dynamically as a function of the stimulation (see \cite{Wannawas2021, Wannawas2023a} for more details). The arm's movement is controlled by applying muscle excitation ($e\in[0,1]$, with 1 representing fully excited) to the stimulated muscles. This excitation can be view as normalised stimulation current in real-world settings. The excitation causes the muscle to activate following a first-order excitation-activation dynamics: $\frac{da_m}{dt}=\frac{1}{\tau} (e-a_m)$, where $a_m$ and $\tau$ are the muscle activation and time constant, respectively. We change the time constant from the default value of $40$ ms to $100$ ms to represent the longer delay of FES-induced activation \cite{Kralj1973}. The observable environment states are the angle and angular velocities of the shoulder and elbow ($\textbf{o}_t = [\boldsymbol{\theta}_{s,t}; \boldsymbol{\theta}_{e,t}; \boldsymbol{\dot{\theta}}_{s,t}; \boldsymbol{\dot{\theta}}_{e,t}]^T$).

\emph{The RL algorithm} of choice is soft actor-critic \cite{Haarnoja2019_1}. Both actor and critic are parameterised by fully-connected neural networks with two hidden layers. The actor's output layer has a sigmoid activation function to squash the outputs within $[0,1]$.

\emph{The RL task here} is to apply the muscle stimulation to move the arm to the desired poses which are specified by target joint angles--shoulder and elbow ($\boldsymbol{\theta}_{tar,t}$). The state vector $\textbf{s}_t$ is $[\textbf{o}_t; \boldsymbol{x}_t; \boldsymbol{\theta}_{tar,t}]^T$. The action vector $\textbf{a}_t$ ($a_{t_i}\in[0,1]$) comprises muscle excitation of the stimulated muscles. The immediate reward $r_t$ is simply computed using the square error and action penalty as $r_t = -(\boldsymbol{\theta}_t - \boldsymbol{\theta}_{tar,t})^2 - (\frac{1}{n}\Sigma_{i=0}^n a_i)$, where $n$ is the number of stimulated muscles.

\emph{The training} is episodic. Each episode has 100 time steps with a 100 ms time step size. The episodes begin at random poses, targets, and fatigue levels. A new random target is assigned at the 50\textsuperscript{th} time step. Every 5 training episodes, the control performances are evaluated in RMSE on 50 test episodes with the same settings as the training episodes.

\section{Results}
\subsection{Ensemble transition function}
\begin{figure}[h!]
    \begin{center}
    \includegraphics[width=0.75\linewidth]{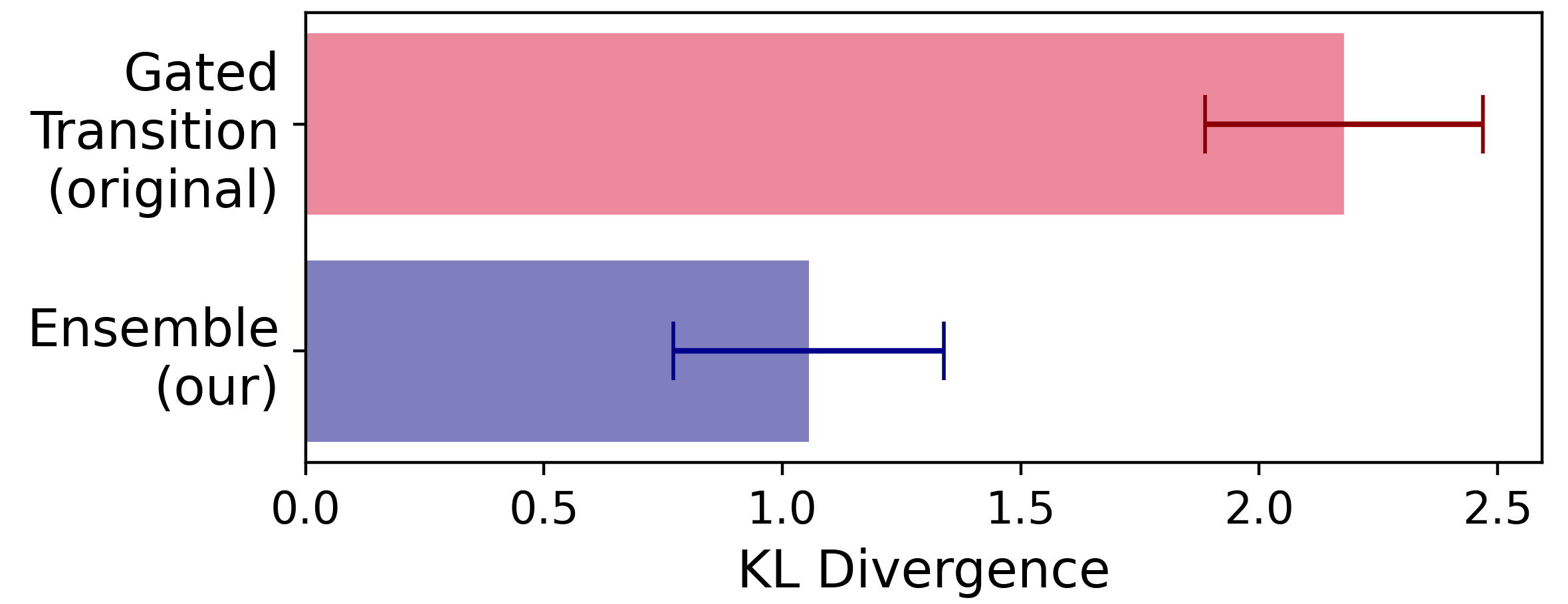}
    \vspace{-5pt}
    \caption{KL Divergences of \textcolor{red}{(red)} the original and (\textcolor{blue}{blue}) our ensemble models measured against the Gaussian Process fitted to the ground truth transition data. The error bars show the standard deviations across 10 runs.}
    \label{fig:KLD}
    \end{center}
    \vspace{-15pt}
\end{figure}

We replace $f_{\mbox{Trans}}$ of the original GSSM model \cite{Krishnan2017} called Gated Transition Function, denoted as $f_{Tr,Ori}$, with RP-Ensemble, denoted as $f_{Tr,Ens}$, to address the overconfidence issue. We test both models on a benchmarking function--Kink \cite{Ialongo2019}--whose ground truth transition is known. We fit the ground truth transition data using a Gaussian Process (GP), a gold standard for uncertainty quantification, against which the learned transitions of both $f_{Tr,Ori}$ and $f_{Tr,Ens}$ are evaluated in KL Divergence measure. Fig.\ref{fig:KLD} shows that $f_{Tr,Ens}$'s KL Divergence with respect to the GP (blue) is lower than that of $f_{Tr,Ori}$ (red). This means that the distributions of the learned transition functions obtained from our ensemble model are closer to the gold standard. This implies that the ensemble model ($f_{Tr,Ens}$) can help improve GSSM and yields better transition function ($f_{\mbox{Trans}}$).

\begin{figure}[h!]
    \begin{center}
    \includegraphics[width=0.91\linewidth]{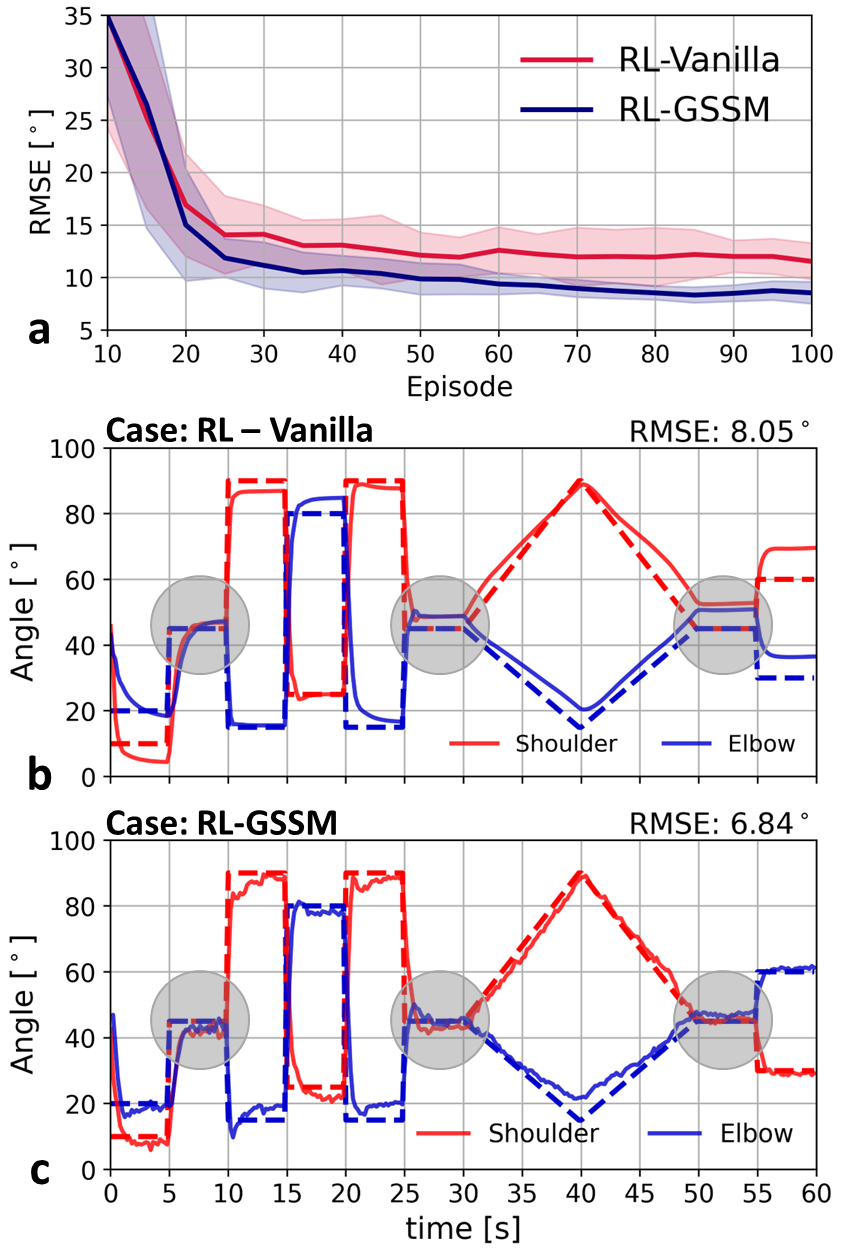}
    \vspace{-10pt}
    \caption{(a) The control performances evaluated along the training. The shades show the standard deviations of 10 runs. (b and c) Control behaviours of (b) $RL-vanilla$ and (c) $RL-GSSM$ in tracking an arbitrary target trajectory. The dash and solid lines represent the target and actual angles, respectively.}
    \label{fig:result_tracking}
    \end{center}
    \vspace{-15pt}
\end{figure}

\subsection{Controlling planar arm movements}
We train our RL-GSSM to control planar arm movements through muscle stimulation under progressive muscular fatigue. We explore 2 cases: the 1) \emph{RL-vanilla} case where the fatigue is unobservable; and 2) \emph{RL-GSSM} case where the GSSM with ensemble transition function is applied. The RL agents are trained for 100 episodes in both cases; the training is repeated 10 times. Fig.\ref{fig:result_tracking}a shows the performance evaluations in RMSE along the training. \emph{RL-vanilla}'s has slightly steeper learning curve at the beginning but stagnates at worse levels. \emph{RL-GSSM}'s curve has higher standard deviations in the early period because the agents have to simultaneously learn the controls and follow the not-yet-converged GSSM. \emph{RL-GSSM}'s performance continuously improves to saturated levels slightly below $10^\circ$ with the smaller variations, which imply higher consistency than those of \emph{RL-vanilla}.

Fig.\ref{fig:result_tracking}b and c show example behaviours in tracking an arbitrary trajectory. Both \emph{RL-vanilla} and \emph{RL-GSSM} can produce good tracking along the 60-second trial. \emph{RL-vanilla}'s performance (Fig.\ref{fig:result_tracking}b solid lines) gradually decreases as the trial progresses with the increasing muscular fatigue. The grey circles highlight noticeable comparison points. \emph{RL-GSSM} (Fig.\ref{fig:result_tracking}c) can bring the shoulder and elbow angles to the $[45^\circ,45^\circ]$ targets at anytime when requested. \emph{RL-vanilla} (Fig.\ref{fig:result_tracking}b), however, tends to lose its performance in the second and third requests as the actual angles increasingly deviate from the targets. The \emph{RL-GSSM}'s overall performance in RMSE along the trajectory is $6.84^\circ$, which is slightly better than that of \emph{RL-vanilla}. The performance is at a comparable level to those in \cite{Crowder2021,Abreu2022} where the targets with the radius size of $5$ cm can be accurately reached. Note, however, that \cite{Crowder2021,Abreu2022} use different arm models which might affect the performance results.


\section{Discussions \& Conclusions}
We present an AI-based approach for controlling FES under progressive muscular fatigue. Our \emph{RL-GSSM} approach uses RL to learn the control policies and GSSM, which is modified to address the overconfidence issue, to provide Markovian state representations to the RL. We demonstrate our approach in controlling arbitrary planar arm movements using a detailed neuromechanical model and compare its control performance to that of \emph{RL-vanilla}--the simple setting that uses a typical RL straightaway. We show that our \emph{RL-GSSM}, unlike \emph{RL-vanilla}, can maintain its control performance against progressive fatigue, yielding higher and more consistent control accuracy.

Several improvements could be added to push \emph{RL-GSSM} toward real-world applications. A simple one is to incorporate the arm model into RL training to reduce the amount of data that needs to be collected from real human-in-the-loop systems. Regarding \emph{RL-GSSM} system itself, the training of RL and GSSM is currently disconnected. Joining the RL's and GSSM's training, for example, through a single cost function could benefit the overall process. Dynamic AI-based adaptation to fatigue or habituation of stimuli via our \emph{RL-GSSM} method will benefit not only muscle-based applications but can be generalised to many other closed-loop neurostimulation or neuromodulation applications (e.g. \cite{lorenz2016automatic,scangos2021closed}).

\bibliographystyle{IEEEtran}
\bibliography{bib}

\end{document}